\documentclass[pra,showpacs,twocolumn,floatfix,superscriptaddress]{revtex4}
\usepackage{graphicx,amsmath,amssymb,subfigure}
\usepackage{epsf,color}
\usepackage{floatflt,latexsym}

\newcommand{\rme}{\text{e}}
\begin{document}
\title{Resolution and sensitivity of a Fabry-Perot interferometer with a
  photon-number-resolving detector}  
\author{Christoph F. Wildfeuer}
\email{wildfeuer@phys.lsu.edu}
\affiliation{Hearne Institute for Theoretical Physics, Department of Physics
  and Astronomy, Louisiana State University,
 Baton Rouge, Louisiana 70803}
\author{Aaron J. Pearlman}
\affiliation{Optical Technology Division, National Institute of Standards and
  Technology, 100 Bureau Drive, Gaithersburg, MD 20899-8441}
\author{Jun Chen}\author{Jingyun Fan}\author{Alan Migdall}
\affiliation{Optical Technology Division, National Institute of Standards and
  Technology, 100 Bureau Drive, Gaithersburg, MD 20899-8441}
\affiliation{Joint Quantum Institute, University of Maryland, College Park, MD 20742}
\author{Jonathan P. Dowling}
\affiliation{Hearne Institute for Theoretical Physics, Department of Physics
  and Astronomy, Louisiana State University,
 Baton Rouge, Louisiana 70803}
\begin{abstract}
With photon-number resolving detectors, we show compression of interference fringes with increasing photon numbers
  for a Fabry-P\'erot interferometer. This feature provides a higher precision
  in determining the position of the interference maxima compared to a
  classical detection strategy. We also theoretically show
  supersensitivity if $N$-photon states are sent into the interferometer and a
  photon-number resolving measurement is performed.  
\end{abstract}
\maketitle
\section{Introduction}\label{Intro}
\par Interferometers with coherent light are one of the building
blocks for high-precision metrology. Recent progress in the field of
photon-number resolving detectors has made
it possible to explicitly measure the photon statistics of different quantum-light sources
in interferometric schemes \cite{Lincoln,Miller,Waks}. One such detector, the
transition edge sensor (TES), is a superconducting micro-bolometer that has
demonstrated very high detection efficiency (95\% at $\lambda=1550\,\mathrm{nm}$) and high photon
number resolution \cite{Appl,Woo,Lita}.  
\par Expanding the average intensity of an
interference pattern into its
photon-number resolved components provides a better understanding of the
interplay of sensitivity and resolution of an interferometer. Using a TES, we are now able
to observe photon-number resolved interference fringes and learn how they
differ from a classical photon-averaged signal. Although it is also possible to
obtain the photon-number resolved interference fringes with multiplexed single photon
counter modules \cite{Mitchell}, it is advantageous to
use a photon-number resolving detector like a TES which provides a high detection
efficiency. The TES offers the advantage of high fidelity detection (high
probability of detecting the correct number of incident photons). On the other
hand, photon-number resolving configurations that rely on multiplexed
single-photon counters, which although have made exceptional progress in
recent years, still suffer from limited photon-number resolving fidelity
\cite{Ralph,Dauler}.
\par For stand-off applications, such as a laser ranging device, it is typical to use coherent states, since they are more robust under loss than
nonclassical states of light. A known strategy to improve the sensitivity of
an interferometer is to squeeze the vacuum of the unused
port of an interferometer, which was first demonstrated by Caves \cite{Caves}. 
Another promising strategy for
quantum sensors is to maintain a coherent laser light source, but
replace the classical intensity measurement with a photon-number resolving detector, or
employ other more
complicated entangling measurements to improve the performance of the quantum
sensor further \cite{Resch}. In addition, the performance of different
non-classical input states, together with a photon-number resolving detection
scheme, may be investigated. Under some conditions, the resolution and, in
particular cases, the sensitivity of these quantum sensors may exceed the performance of `classical' light
sources and detection schemes. We emphasize here and highlight later that
while resolution and sensitivity are related, they are not identical. For an overview of quantum metrology
applications, see Ref.~\cite{Hwang}. 
\par Many authors have proposed resolution and
sensitivity enhancements in different types of interferometric schemes, where a large
variety of Sagnac, Michelson, Mach-Zehnder, and Fabry-P\'erot interferometers (FPI)
are considered
\cite{Boto,Boyd,Shapiro,Giovannetti,Tsang,Lloyd,Khoury,Agarwal,Huver}. A notable
example is the laser interferometer gravitational wave
observatory (LIGO) that consists of a Michelson interferometer with
Fabry-P\'erot cavities in
each of the two arms to boost the overall sensitivity of the device \cite{Kimble}.
\par We first theoretically investigate the photon-number resolved interference fringes of a
Fabry-P\'erot interferometer as we scan its phase. We then experimentally implement
this with a TES. Similarly, Khoury {\em et al.} have reported the use of a visible
light photon counter to monitor the output of a Mach-Zehnder interferometer \cite{Khoury}. We compare resolution
and sensitivity of the photon-number resolved interference pattern with the
classical case where a coherent state is sent through a FPI and only the
average intensity is measured. We also theoretically investigate the performance of
resolution and sensitivity for a single-mode photon-number state $|n\rangle$ in
combination with a photon-number resolving detection of the interference fringes.
\section{Quantized description of a Fabry-P\'erot interferometer}
\par
\begin{figure}[htb]
\includegraphics[scale=0.75]{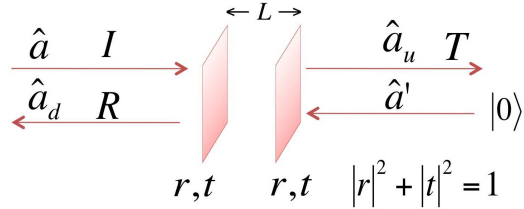}
\caption{(Color online) Fabry-P\'erot cavity with complex amplitudes $R$
  and $T$ for reflected and transmitted modes, respectively, and incident
  intensity $I$. Each
  mode can be assigned a mode operator marked by the hat to quantize the
  respective mode, where the subscripts $u$ and $d$ stand for up and down. We
  assume, for simplicity, that both mirrors have identical complex reflection and transmission
  coefficients denoted with $r$ and $t$.\label{FPI1}}
\end{figure}

 We start our investigation by deriving a quantum mechanical description for the
  Fabry-P\'erot interferometer. Loudon first considered a quantum theory of the FPI for high-resolution length measurements \cite{Loudon}. The two incoming and two outgoing modes of the FPI can
  be quantized as displayed in Fig.~\ref{FPI1}.
\par
\begin{figure}[htb]
\includegraphics[scale=0.7]{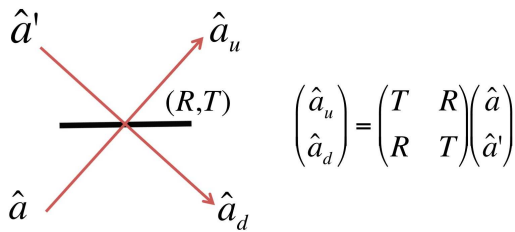}
\caption{(Color online) Effective beam-splitter for the Fabry-P\'erot cavity with complex amplitudes $R$
  and $T$ for reflected and transmitted modes, respectively, which
  satisfy the conditions $|T|^2+|R|^2=1$ as well as $TR^*+RT^*=0$. \label{BS1}}
\end{figure}

 The modes described in Fig.~\ref{FPI1} can be
  transformed by an effective beam-splitter (BS) transformation, as is displayed in Fig.~\ref{BS1}.
For a FPI with two identical highly reflecting mirrors, the
  transmission and reflectance functions $T$ and $R$ are given by \cite{Loudon}
\begin{eqnarray}
  T(r,\phi)&=&\frac{(1-|r|^2)\rme^{-2i\sqrt{1-|r|^2}}}{|r|^2\rme^{-2i\sqrt{1-|r|^2}}\rme^{2i\phi}-1},\label{FPI-functionsa}\\
  R(r,\phi)&=&\frac{|r|\rme^{-i\sqrt{1-|r|^2}}(\rme^{-i\phi}-\rme^{i\phi}\rme^{-2i\sqrt{1-|r|^2}})}{|r|^2\rme^{-2i\sqrt{1-|r|^2}}\rme^{2i\phi}-1},\label{FPI-functionsb}
\end{eqnarray}
where $r$ denotes the complex reflectivity of the mirrors, and $\phi=kL=2\pi L/\lambda$
denotes a phase determined by the wave-number $k$ of the incoming light,
and the distance $L$ between the two mirrors \cite{Loudon}.
\par As a classical baseline, we consider a single-mode coherent state given
by \cite{Glauber}
\begin{equation}
|\alpha\rangle=\mathrm{e}^{-\frac{|\alpha|^2}{2}}\sum_{k=0}^\infty
 \frac{\alpha^k}{\sqrt{k!}}|k\rangle\,,
\end{equation} 
(where $|k\rangle$ is a $k$-photon Fock state and $\alpha$ is the dimensionless
electric field amplitude of the coherent state with the mean photon number $\bar{n}=|\alpha|^2$), which describes very well a single-mode laser above threshold. This state is incident on the FPI in mode $\hat{a}$, and vacuum
$|0\rangle$ goes in mode $\hat{a}'$. The two-mode input state $|\alpha\rangle_{\hat{a}}|0\rangle_{\hat{a}'}=|\alpha,0\rangle_{\hat{a}{\hat{a}'}}$ is then
transformed with the BS transformation in Fig.~\ref{BS1}, from which we
obtain $\hat{a}^\dagger=T\hat{a}_u^\dagger+R\hat{a}^\dagger_d$, where the
subscripts {\it u} and {\it d} stand for up and down, respectively. Note that $T$ and $R$ satisfy the conditions $|T|^2+|R|^2=1$ as well as $TR^*+RT^*=0$. 
\par
\begin{figure}[htb]
\subfigure[]{\label{classical0}\includegraphics[scale=0.6]{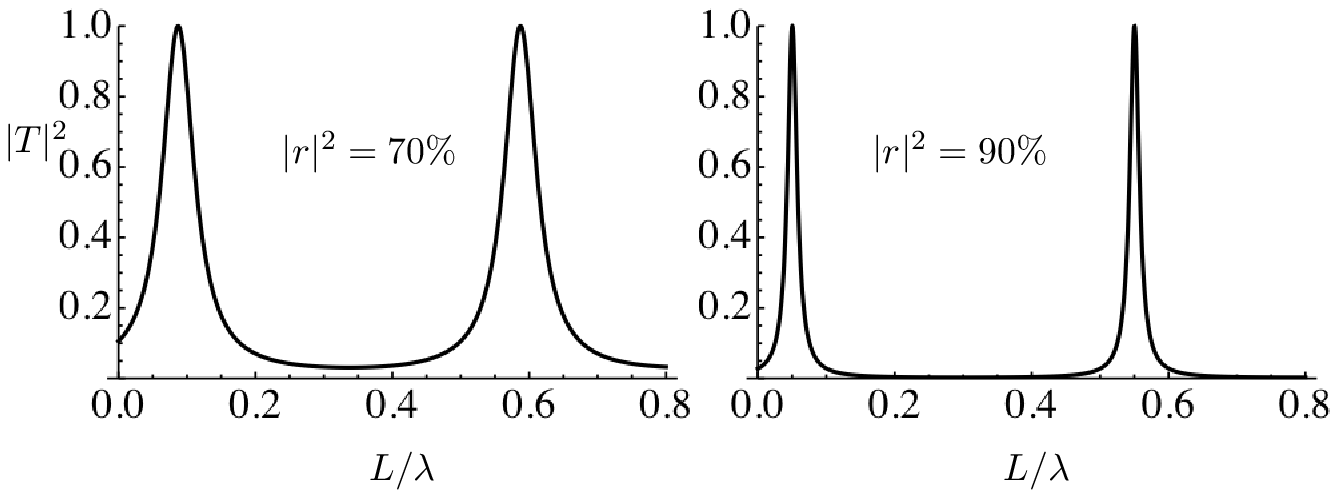}}
\subfigure[]{\label{pnrcoherentandfock}\includegraphics[scale=0.6]{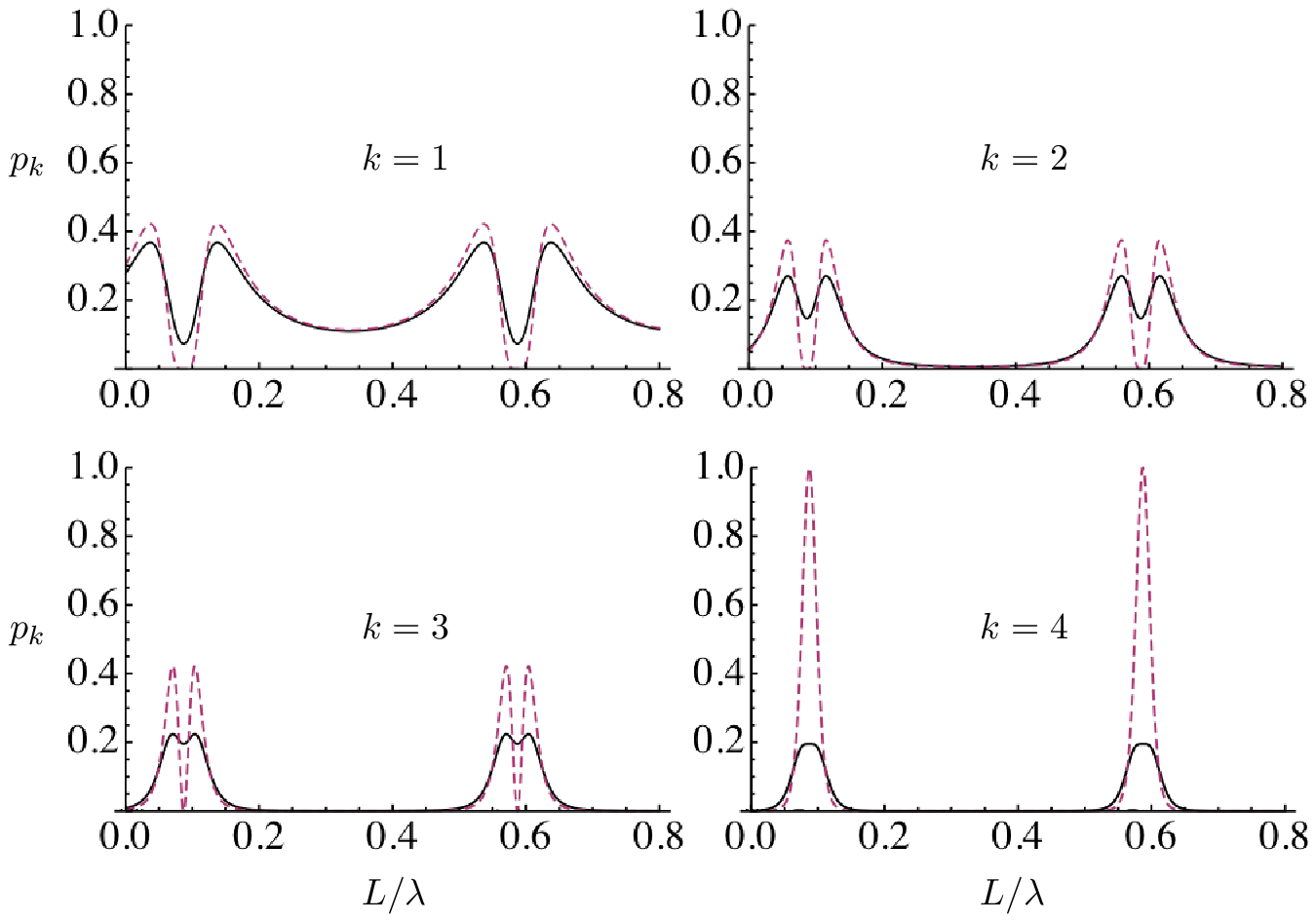}}
\caption{(a) shows the classical transmission function $|T|^2$ displayed for a reflectivity of
  $|r|^2=70\%$ (LHS) and $|r|^2=90\%$ (RHS) of the FPI-mirrors. Note that the
  transmission peaks become
     narrower and the maxima slightly shifted to the left as the reflectivity
  increases. Also the minima go to zero for $90\%$ reflectivity whereas they
  do not reach zero for lower reflectivities. (b) shows the probabilities of
  detecting $k$ photons, $p_k^\mathrm{coh}$ for a single-mode coherent input state $|\alpha\rangle$
  with mean photon number $\bar{n}=|\alpha|^2=4$ incident on the FPI and a photon-number resolving
  measurement displaying $k=1,..,4$ (solid lines). The 
  reflectivity $|r|^2$ of the mirrors is $70\%$. The dashed lines shows the transmission
  probabilities $p_k^\mathrm{F}$ for a single-mode photon-number state
  $|4\rangle$. We observe that the Fock states (dashed lines) show transmission peaks
  that are sharper in general. This effect is equivalent to operating at a
  larger reflectivity of the mirrors, which is demonstrated in (a) for the classical
  curves. A major difference appears for $k=4$ where the transmission maxima
  reach one for the Fock state, whereas the coherent state stays
  low. Here and in the following, we choose a reflectivity of
  $70\%$ as the important features are more pronounced than for larger reflectivities.}
\end{figure}

 We transform the incident coherent state
$|\alpha,0\rangle_{\hat{a},\hat{a}'}$ by the effective BS transformation and
obtain the output of the FPI. An ideal $k$-photon
detection is described by the projector $\hat{C}=|k\rangle\langle
k|$. An approach including detection efficiencies is presented in
Ref.~\cite{Hwang2}. Applying this on mode $\hat{a}_u$, we
obtain the photon-number resolved interference fringes, which leads to the
expression for the probability of detecting $k$ photons
\begin{eqnarray}\label{coherent-state-k}
  \lefteqn{
  p_k^{\mathrm{coh}}=\mathrm{Tr}\left(\hat{C}\hat{\rho}^\mathrm{coh}\right)}\nonumber\\
  &&=\mathrm{e}^{-\bar{n}}\sum_{j=k}^\infty
  \frac{\bar{n}^j}{k!(j-k)!}|T|^{2k}(1-|T|^2)^{j-k},
\end{eqnarray} 
where $\hat{\rho}^\mathrm{coh}$ is the reduced density matrix for the coherent
state in mode $\hat{a}_u$. This result is displayed
in Fig.~\ref{pnrcoherentandfock} for a mean photon number of four
($\bar{n}=4$) and $k=1,..,4$ detected photons, which shows $p_k^\mathrm{coh}$
as a function of $L/\lambda$ (or $\phi/2\pi$).
 \par The structure of the transmission functions may be understood from
interpreting the terms in Eq.~(\ref{coherent-state-k}). Each term in the sum
represents the probability that $k$ photons are transmitted through the FPI,
multiplied by the probability that $j-k$ photons are reflected off the
FPI. The transmission probabilities $|T|^{2k}$ have a maximum where the reflection probability $(1-|T|^2)^{j-k}$ has its
minimum. The multiplication
of both probabilities results in the additional minimum in the transmission
probabilities for $k< \bar{n}$. 
\par We can also calculate the 
response of a non-photon-number resolving detector by calculating the
expectation value
\begin{equation}\label{classical}
  p^\mathrm{coh}=\langle \hat{a}^\dagger_u \hat{a}_u\rangle^\mathrm{coh}=\sum_{k=1}^\infty k\, p_k^\mathrm{coh}=\bar{n}|T|^2\,,
\end{equation}
which is proportional to the mean number of photons
incident on the detector.
We refer to this result as the `classical' signal that is usually associated
with the output of a FPI. 
\par
\begin{figure}[htb]
\includegraphics[scale=0.8]{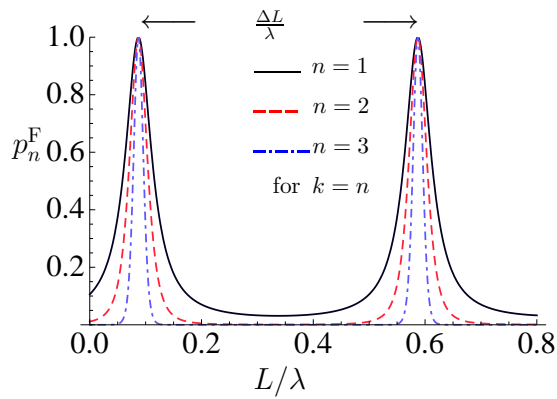}
\caption{(Color online) Transmission probabilities for a single-mode photon-number
  state and a photon-number resolving measurement, for photon numbers
  $n=1,2,3$. The free spectral range $\Delta L/\lambda$ is the distance
  between two adjacent maxima. Transmission probabilities are calculated for
  70\% reflectivity of the mirrors. \label{FPI-output}}
\end{figure}

 For a nonclassical input we consider a single-mode Fock state $|n\rangle$ in mode
$\hat{a}$ incident on the FPI, and the vacuum state $|0\rangle$ in mode $\hat{a}'$.  
The input state $|n\rangle_{\hat{a}}|0\rangle_{\hat{a}'}=|n,0\rangle_{\hat{a},\hat{a}'}$ is then transformed to
\begin{eqnarray}
 \lefteqn{|n,0\rangle_{\hat{a},\hat{a}'}=\frac{(\hat{a}^\dagger)^n}{\sqrt{n!}}|0,0\rangle_{\hat{a},\hat{a}'}\rightarrow\frac{(T\hat{a}_u^\dagger+R\hat{a}_d^\dagger)^n}{\sqrt{n!}}|0,0\rangle_{{\hat{a}_u},{\hat{a}_d}}}\nonumber\\
 &&=\frac{1}{\sqrt{n!}}\sum_{\ell=0}^n {n\choose\ell}T^\ell R^{n-\ell}\sqrt{\ell!(n-\ell)!}|\ell,n-\ell\rangle_{{\hat{a}_u},{\hat{a}_d}}.\nonumber\\
\end{eqnarray}
Suppose we also perform a photon-number resolving measurement on the
transmitted photons. The
result of this measurement is given by 
\begin{equation}\label{Fock-transmission}
p_k^\mathrm{F}=\mathrm{Tr}\left(\hat{C}\hat{\rho}^\mathrm{F}\right)=\frac{n!}{k!(n-k)!}|T|^{2k}(1-|T|^2)^{n-k}\,,
\end{equation}
where $\hat{\rho}^\mathrm{F}$ is the reduced density matrix for the Fock state
in mode $\hat{a}_u$ and the superscript F denotes Fock state. Note that the
right hand side of Eq.~(\ref{Fock-transmission}) is a binomial distribution
with the property $\sum_{k=0}^n p_k^\mathrm{F}=1$.
The detection probabilities $p_k^\mathrm{F}$ in Eq.~(\ref{Fock-transmission})
are displayed in Fig.~\ref{pnrcoherentandfock} (dashed lines) as a function of
$L/\lambda$ for a four-photon state and a photon-number detection for $k=1,\ldots,4$. 
The mean photon counts at the output of the FPI are obtained from the expectation value
\begin{equation}\label{Fock-transmission2}
  \langle\hat{a}^\dagger_u\hat{a}_u\rangle^\mathrm{F}=\sum_{k=1}^n\,k\, p_k^\mathrm{F}=n|T|^{2}\,.
\end{equation}
The result of Eq.~(\ref{Fock-transmission2}) is the same as in the classical
case (Eq.~(\ref{classical})) above, when we identify $\bar{n}=n$. However, the
photon-number resolving measurements in Eq.~(\ref{Fock-transmission}) show a
different behavior. In particular, we consider the case $k=n$, i.e., we
measure the same photon number in our detector as that of the initial input
state. 
In this case Eq.~(\ref{Fock-transmission})
reduces to 
\begin{equation}
  p_n^\mathrm{F}=|T|^{2n}\,.
\end{equation}
It turns out that the transmission peaks become narrower as the
photon number $n$ increases (Fig.~\ref{FPI-output}).
We also observe a similar interference pattern for the photon-number
resolved peaks for $k<n$ for the same reason as described earlier for coherent
states. As in the coherent case, for $k<n$ the transmission probability
$|T|^{2k}$ for $k$ photons is multiplied by the
probability $(1-|T|^2)^{n-k}$ which is the probability that the other $n-k$
photons are reflected. The multiplication of these two probabilities, which
have opposite functional forms, produces the dip in the middle of
the maximum.  
\section{Resolution and sensitivity for coherent states}
\par To quantify our results for different states and detection
operators, we first calculate the uncertainty in determining
the free spectral range (FSR) given by $\Delta L/\lambda$, which is the
dimensionless distance between two adjacent interference peaks (Fig.~\ref{FPI-output}).
From the experimental data we can determine the variance of
the transmission peaks. The uncertainty of the absolute positions, or in other
words, the standard deviation of the mean (SDM) value of the individual
peaks one and two are given by $\sigma_{L_1}=\sigma_1/\sqrt{n_1}$, and
$\sigma_{L_2}=\sigma_2/\sqrt{n_2}$, where $n_1$ and $n_2$ are the total number of
counts in peak one and two. For an explicit definition of the SDM $\sigma_{L_i}$ see section \ref{experiment}, which
contains our experimental results. The uncertainty in determining the
FSR is then given by
\begin{equation}\label{sigma_L}
  \sigma_{\Delta
  L}=\sqrt{\frac{\sigma_1^2}{n_1}+\frac{\sigma_2^2}{n_2}}\approx\sqrt{2}\frac{\sigma}{\sqrt{n}}\,,
\end{equation} 
where the approximation holds, if the two peaks have approximately the same variance and
number of counts. A smaller variance $\sigma$ for the transmission peaks provides an
improvement in resolution, but increasing the number of counts can provide a
similar improvement. If the variance shrinks by a factor of $m$, i.e., 
$\sigma\rightarrow \sigma/m$, the uncertainty of the peak center in
Eq.~(\ref{sigma_L}) becomes
$\sqrt{2}\sigma/(m^2 n)^{1/2}$. This means
that we need a factor of $m^2$ fewer counts to obtain the original variance $\sigma$. 
In section \ref{experiment} we show experimentally that we can determine the position of the peaks from
the photon-number resolved data obtained with coherent states with up to three times
higher precision, for the same optical power compared to the classical signal. 
\par On the other hand, we
can compute the uncertainty $\delta L$ of a length measurement, which we also refer to as sensitivity, from the expression  \begin{equation}\label{uncertainty}
  \delta
  L=\frac{\Delta\hat{C}}{\left|\partial\langle\hat{C}\rangle/\partial L\right|}\,,
\end{equation}
where $\langle \hat{C}\rangle$ is the mean value of the detection
operator and
$\Delta\hat{C}=(\langle\hat{C}^2\rangle-\langle\hat{C}\rangle^2)^{1/2}$ is the
standard deviation of the observable $\hat{C}$ \cite{Dowling}. 
\par
\begin{figure}[htb]
\includegraphics[scale=0.65]{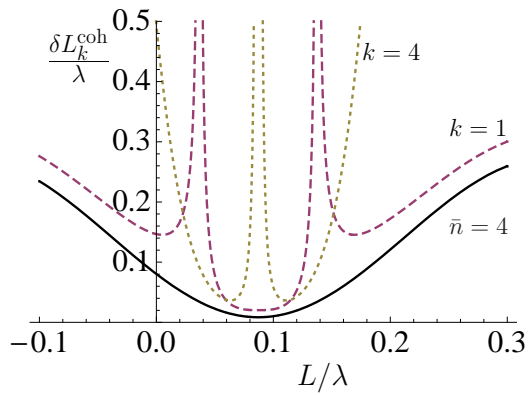}
\caption{(Color online) Dimensionless uncertainty $\delta L^\mathrm{coh}_k/\lambda$ (where
  $L$ is the length change of the FPI and $\lambda$ is the wavelength of the
  coherent laser beam) for a mean intensity measurement with $\bar{n}=4$
  compared to photon number resolving measurements $k=1$
  and $k=4$. The solid line also represents the shot-noise limit. The
  reflectivity $|r|^2$ of the mirrors is $70\%$.\label{photon-number-resolved}}
\end{figure}

 Next we compute the sensitivity $\delta L^\mathrm{coh}_{\bar{n}}$ for a coherent state with
average photon number $\bar{n}$ as a classical baseline, and compare
it with the photon-number resolved transmission probabilities (Fig.~\ref{photon-number-resolved}).
For a coherent state input and a mean intensity measurement,
Eq.~(\ref{uncertainty}) reduces to
\begin {equation}\label{coherentaverage}
  \delta L^\mathrm{coh}_{\bar{n}}=\frac{1}{\sqrt{\bar{n}}}\frac{|T|}{|\partial
  |T|^2/\partial L|}\,.
\end{equation}
This expression defines the shot-noise limit evidenced by the
$1/\sqrt{\bar{n}}$ dependence. Improving the sensitivity beyond this is referred to as supersensitivity.
For the photon-number resolved sensitivity $\delta L_k^\mathrm{coh}$,
Eq.~(\ref{uncertainty}) reduces to
\begin{equation}
  \delta L_k^\mathrm{coh}=\frac{\sqrt{p_k^\mathrm{coh}(1-p_k^\mathrm{coh})}}{|\partial
  p_k^\mathrm{coh}/\partial L|}\,.
\end{equation}
We observe that the uncertainty of a length measurement $\delta L^\mathrm{coh}/\lambda$ for
the photon-number resolved measurement is always larger than for the average
photon-number measurement. In other words the photon-number resolved
interference fringes do not increase the sensitivity of the
interferometer. However, we show in the next section that increased sensitivity can be achieved
by replacing the input coherent state with a photon-number state.  
\section{Sensitivity for $N$-photon states}\label{section-Fock}
\par
\begin{figure}[htb]
\includegraphics[scale=0.5]{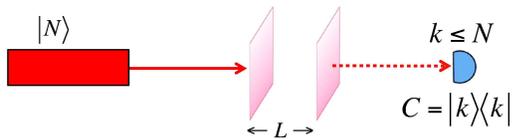}
\caption{(Color online) Transmission experiment with an $N$-photon state
  incident on a Fabry-P\'erot interferometer and photon-number resolving
  detection of the output.\label{N-photon-transmission}}
\end{figure}

\begin{figure}[htb]
\includegraphics[scale=0.55]{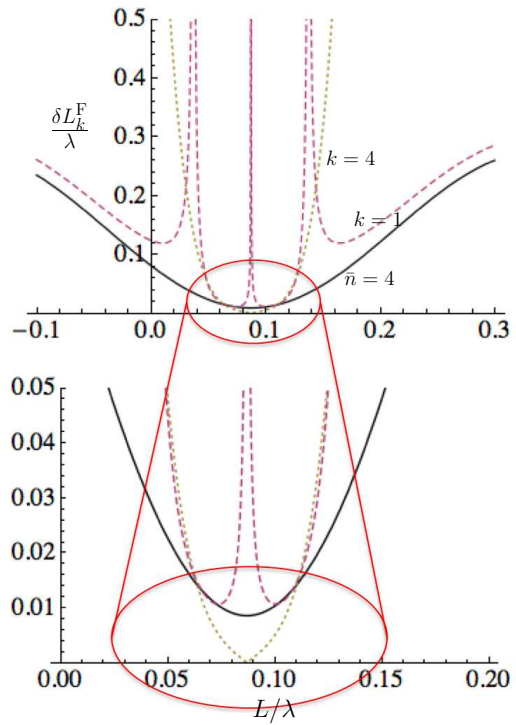}
\caption{(Color online) Dimensionless uncertainty $\delta L^\mathrm{F}_k/\lambda$ (where
  $L$ is the length change of the FPI and $\lambda$ is the wavelength of the light) for a mean intensity measurement with $\bar{n}=4$
  compared to photon number resolving measurements $k=1$
  and $k=4$. The solid line also represents the shot-noise limit. The
  reflectivity $|r|^2$ of the mirrors is $70\%$. The shot-noise limit (solid line) is beaten by the $k=4$ curve.\label{kFock1}}
\end{figure}

 We propose an experiment with an $N$-photon state $|N\rangle$ incident on the
FPI as displayed in Fig.~\ref{N-photon-transmission}. 
From the Fig.~\ref{pnrcoherentandfock} (dashed line for $k=4$ ), and Fig.~\ref{FPI-output} we conclude that in addition to obtaining narrower transmission
functions as the photon-number increases (increased resolution), the amplitude at the maximum
remains one. That is an indication
that we have an additional benefit from using $N$-photon states as opposed to
a coherent state input. We not only increase the resolution, we also obtain a
higher sensitivity. To quantify this statement we calculate the 
sensitivity as defined by Eq.~(\ref{uncertainty}).  
The sensitivity is given by 
\begin{equation}
  \delta L_k^\mathrm{F}=\frac{\sqrt{p_k^\mathrm{F}(1-p_k^\mathrm{F})}}{|\partial
  p_k^\mathrm{F}/\partial L|}\,,
\end{equation}
where $p_k^\mathrm{F}$ is taken from the expression in
Eq.~(\ref{Fock-transmission}), which simplifies for a $|k\rangle$ Fock state input
and a $k$-photon detection to
\begin{equation}\label{focknumberresolved}
  \delta L^\mathrm{F}_k=\frac{|T|^k\sqrt{1-|T|^{2k}}}{|\partial |T|^{2k}/\partial L|}\,.
\end{equation} 
We display the sensitivity as a function of
phase (here scaled length $L/\lambda$) in Fig.~\ref{kFock1}, analogous to the
coherent state case in Fig.~\ref{photon-number-resolved}. 
We observe that the
  shot-noise limit given by the black solid line is beaten by the $k=4$ curve.
  This means that we can achieve supersensitivity (beating the shot-noise
  limit) with a photon-number state input $|n\rangle$ and a $n$ photon
  detection. 
\par
\begin{figure}[htb]
\includegraphics[scale=0.8]{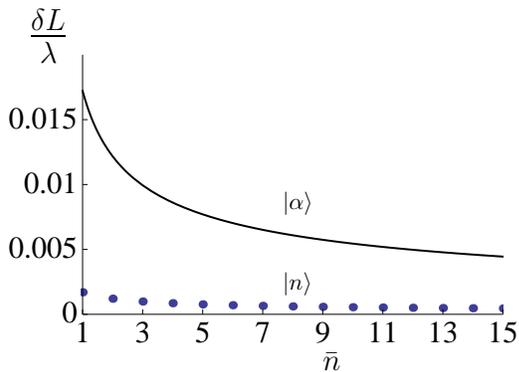}
\caption{(Color online) Comparison of the sensitivity $\delta
  L/\lambda$ (Eq.~(\ref{coherentaverage})) for a
  coherent state (solid line) and mean intensity detection (shot-noise limit) as a function of mean number of detected photons
  versus a $n$-photon state $|n\rangle$ with $n$-photon resolving
  measurement (Eq.~(\ref{focknumberresolved})) as a function of the photon
  number, where we take $n=\bar{n}$, slightly away from the
  transmission maximum. The
  reflectivity $|r|^2$ of the mirrors is $70\%$. The parameter of the phase is
  chosen for each $n$,
  $\bar{n}$, respectively, so that the sensitivity is at its minimum.\label{DeltaL}}
\end{figure}

 We can also investigate how the minimum phase uncertainty behaves as a function of
the photon number $n$ for the photon-number state input or mean photon number $\bar{n}$
for coherent states, respectively. We see that (Fig.~\ref{DeltaL}), as
  opposed to coherent states, a length measurement with photon-number states provides us with a much smaller uncertainty
  $\delta L$ in the few photon limit. Hence, the sensitivity of the FPI is
  increased. An alternative way to describe the result can be formulated
  in terms of the finesse ${\mathcal F}=\mathrm{FSR/FWHM}$ of the FPI, where FWHM stands for full width
  at half maximum. The finesse
  $\mathcal{F}$ of the FPI, which for $R>0.5$ can be approximated by
  ${\mathcal F}= \pi R^{1/2}/(1-R)$ for classical detection, is essentially
  improved by using photon-number resolving detection without changing
  the reflectivity of the mirrors, since the FSR remains the same but the FWHM
  becomes narrower with increasing photon number.
\par Finally, we provide an intuitive interpretation of our results. The
  probability that a single photon traverses through a single beam-splitter,
 described by the complex transmittivity $t$, and reflectivity $r$, is just $|t|^2$. If we
 ask for the probability that $n$ photons in a Fock state traverse through the
 BS, we obtain $|t|^{2n}$ (compare with Eq.~(\ref{Fock-transmission})). In our quantum mechanical model for the FPI,
 we use an effective BS transformation where the matrix elements of the
 unitary BS transformation are given by the complex functions $T$
 and $R$ defined in Eqs.~(\ref{FPI-functionsa}) and (\ref{FPI-functionsb}). We observe then the same
 functional behavior as for the regular BS and Fock states. The transmission
 function for the FPI, given the $n$ photons in a Fock state that have traversed the
 FPI, is $|T|^{2n}$, as given by Eq.~(\ref{Fock-transmission}). As $|T|$ becomes
 smaller than one, the probability of transmitting $n$ photons given by
 $|T|^{2n}$ decreases more
 rapidly than that for single photons ($n=1$) or a coherent state, which manifests as
 narrower transmission curves. This feature may find applications in
 interferometry for high-precision length measurements as in LIGO for instance. The quantum light source
 may provide a high sensitivity at a much reduced optical power. The FPI can also be nested in a Michelson or Mach-Zehnder
 interferometer, as has been implemented at LIGO, to boost the sensitivity and
 achieve an even higher resolving power.
\section{Experimental results for coherent states}\label{experiment}
\par
\begin{figure}[htb]
\includegraphics[scale=0.35]{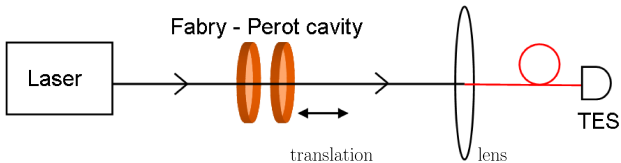}
\caption{(Color online) Transmission experiment with a weak coherent laser
  beam incident on a length tunable FPI and a fiber coupled TES. The light is collected and
  coupled into a single-mode fiber and transmitted to the detector. \label{schematicFPI}}
\end{figure}

 We performed an experiment with an
 attenuated coherent pulsed laser diode at a fixed wavelength $\lambda=1550\,\mathrm{nm}$, a repetition rate of 50 kHz, and a pulse duration
  of 50 ps (Fig.~\ref{schematicFPI}). Our
  photon-number resolving TES detected on average four photons
  per pulse \cite{Woo}. 
We used a scanning Fabry-P\'erot interferometer that had originally been designed as a tunable filter with a FSR of 70 nm
  and a FWHM of 0.15 nm to be used in locked mode. However, its feedback (locking)
  circuit locks only to the maximum of the transmission curve. Since we wanted
  to measure the
  entire transmission function, we had to use the FPI in the unlocked mode. The
  stability of the unlocked FPI was initially poor due to
  ambient temperature fluctuations. To circumvent this, we employed a thermo-electric
  cooler to stabilize the temperature within $0.1\,^{\circ}{\mathrm C}$. For
  the measurement, we attenuated the laser diode output, sent it
  through the FPI and then to the fiber-coupled TES. We adjusted the distance between the mirrors of the FPI by
  tuning the voltage of the piezo-electric transducer inside the FPI.
\par
\begin{figure}[!h]
\includegraphics[scale=0.65]{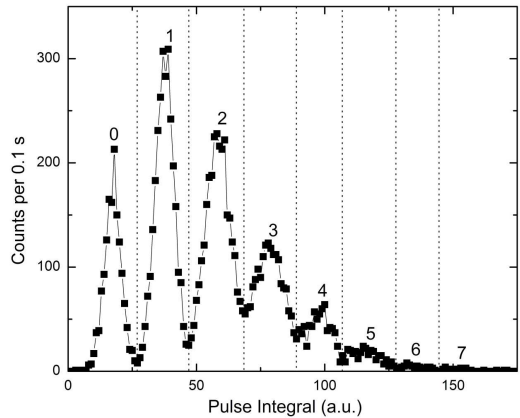}
\caption{Sample histogram of the output pulse integrals of the photon-number resolving TES used in
  our experiment. The histogram indicates the probabilities of detecting each
  photon number, as well as the probability no photons were detected (labeled
 ``0''). The vertical lines show the thresholds between
  the individual photon-number peaks.\label{histogramtes}}
\end{figure}

\begin{figure}[!h]
\includegraphics[scale=0.55]{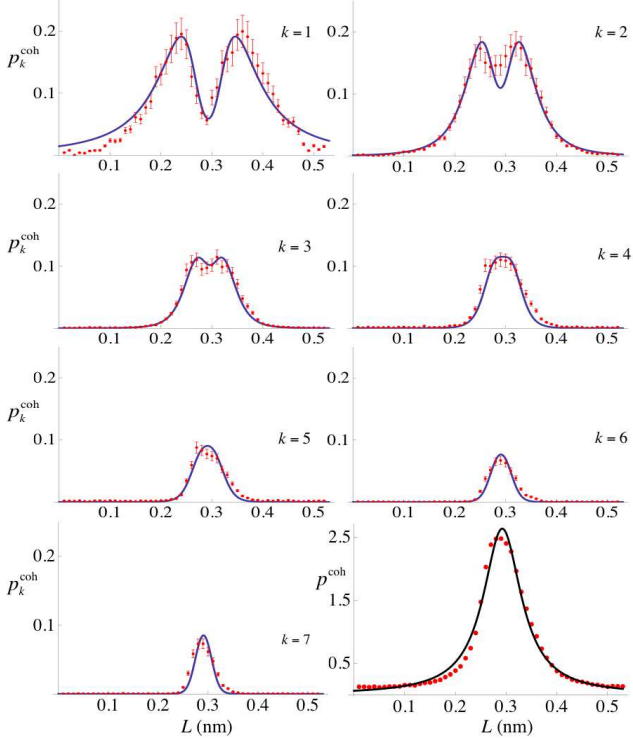}
\caption{(Color online) Transmission probabilities for a single-mode coherent
  state and a photon-number resolving measurement, for photon numbers
  $k=1,..,7$. The blue solid lines show the individual theoretical fits. The
  peaks become narrower as $k$ increases. The fit
  yields the mean photon number $\bar{n}\approx 3.9$ and the
  reflectivity of the mirrors $\approx 91\%$. The graph at the bottom right is the reconstructed classical signal from the photon-number
  resolved curves with $p^\mathrm{coh}=\sum_{k=1}^7 k\,p_k^\mathrm{coh}$. The black solid line is the theoretical fit. From the fit
  the mean photon-number is determined as $\bar{n}'=2.6$, which is not in good
  agreement with the mean photon-number obtained from the photon-number
  resolved curves of $\bar{n}\approx 3.9$ and reflects
  insufficient stability of the used FPI. We also only have access to
  photon-number counts in the range $1\le
  k\le 7$ which amounts to reconstructing the classical signal with an
  amplitude 15\% too low.\label{FPI-output1}}
\end{figure}

 Photon absorption in the TES creates a voltage
pulse whose integral is proportional to the energy absorbed. 
Thus, by simply integrating the output pulses from the TES, we can resolve
straightforwardly the number of photons absorbed in a given time window. In
our setup, we amplify the TES signal and record it using a digital
oscilloscope. We then integrate each pulse and create a histogram of these
pulse integrals to observe the photon-number resolved detection as shown in
Fig.~\ref{histogramtes}. 
We repeat this procedure to obtain a histogram at
each value of the piezo voltage and generate the curves shown in
Fig.~\ref{FPI-output1} for $1\le k\le 7$. The preliminary data support the theoretical predictions, as can be seen in
  Fig.~\ref{FPI-output1}. 
\par Unfortunately, although the stability of the FPI output was improved with
  temperature control, substantial drift occurred during the data acquisition
 time of more than 30 minutes, resulting in smearing of the data.
 Also, the apparent classical signal from the photon-number resolved data is
 systematically underestimated. For weak laser pulses with a mean photon number of four, we
  would need to include photon-number resolved data of up to $k=10$
  to cover $99\%$ of the signal. Our
  measurements only cover $1\le k\le 7$, before the signal disappears into the
 noise, which results in the reconstructed classical signal being 15\% lower than expected.
\par Note that from the theoretical curves shown in
  Fig.~\ref{pnrcoherentandfock}, one can see that there is always a dip in
  the middle of the maximum whenever $k<\bar{n}$. The dip becomes less
  pronounced as $k$ approaches $\bar{n}$, and disappears for $k\ge\bar{n}$. This
  feature can be utilized as a diagnostic tool to bound $\bar{n}$, without
  doing any detailed fitting. For example, from the data in Fig.~\ref{FPI-output1}, we
  can easily identify $3<\bar{n}<4$, just from the ``dip characteristics''
  described above. We can therefore confidently justify the rejection of the
  reconstructed number $\bar{n}'=2.6$ from Fig.~\ref{FPI-output1} (bottom right), which corroborates the
  arguments given in the caption of Fig.~\ref{FPI-output1} and the above text.
\par The photon-number re\-solved output of the FPI (Fig.~\ref{FPI-output1})
  shows narrower peaks with increasing photon number. 
To quantify any improvement in resolution we compare the standard deviations
of the photon-number resolved peaks $\sigma_k$ to the
standard deviation $\sigma_\mathrm{cl}$ obtained from the classical
transmission peak (Table~\ref{Table2}). 
\begin{center}
\begin{table}[!h]
\caption{Resolution improvements. Theo\-reti\-cal and experimental results for the photon-number resolved SDM
  $\sigma^\mathrm{exp}_k$ compared to the SDM $\sigma_\mathrm{cl}^\mathrm{exp}=0.103\,\mathrm{nm}$ of the
  classical signal. The theoretical results are calculated for a mean photon
  number $\bar{n}=3.9$ and $91\%$ mirror reflectivity (fit-parameters determined
  from the experimental data). The expected classical SDM is $\sigma_\mathrm{cl}^\mathrm{theo}=0.0995\,\mathrm{nm}$. All standard
  deviations are given in nm.\label{Table2}}
\begin{tabular}{|c|c|c|c|c|c|c|c|}\hline
k & 1 & 2 & 3 & 4 & 5 & 6 & 7 \\ 
 \hline
$\sigma_k^\mathrm{theo}$ & 0.176 & 0.116 & 0.074 & 0.052 & 0.039 & 0.032 & 0.028 \\
$\sigma_\mathrm{cl}^\mathrm{theo}/\sigma_k^\mathrm{theo}$ & 0.6 & 0.9 & 1.3 & 1.9 & 2.5 & 3.1 &
3.6 \\ \hline 
$\sigma_k^\mathrm{exp}$ & 0.161 & 0.094 & 0.062 & 0.064 & 0.075 & 0.045 & 0.036 \\
$\sigma_\mathrm{cl}^\mathrm{exp}/\sigma_k^\mathrm{exp}$ & 0.6 & 1.1 & 1.7 & 1.6 & 1.4 & 2.3 &
2.9 \\ \hline
\end{tabular}
\end{table}
\end{center}
\begin{figure}[htb]
\includegraphics[scale=0.5]{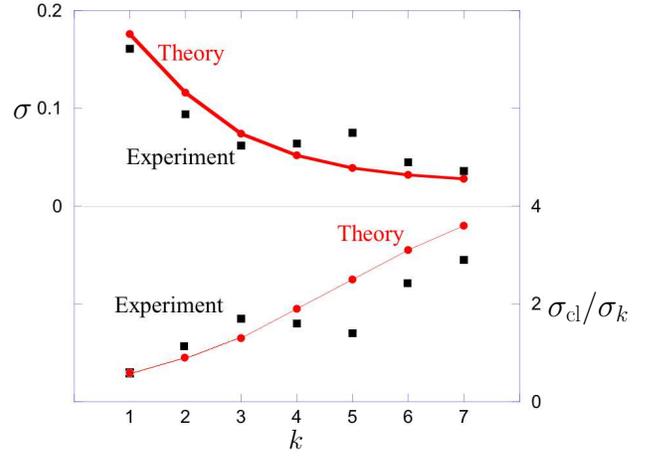}
\caption{(Color online) Figure of the standard deviations as a function of $k$
  presented in Table \ref{Table2}.\label{table_figure}}
\end{figure} 
\par The SDM is defined
by $\sigma^2=\sum_{i}p_i(\phi_i-\mu)^2$, where the mean $\mu$ obtained from
$\mu=\sum_i p_i \phi_i$, $p_i$ is the normalized probability defined by $p_i=f_i/N$, $f_i$ is the number of
counts for a particular phase $\phi_i=2\pi L_i/\lambda$, and $N$ is the total number
of counts for the $k$th transmission peak $\sigma_k$ or the classical signal $\sigma_\mathrm{cl}$. 
For the photon-number
resolved peaks we observe that the SDM gets
smaller for larger photon numbers. This feature allows us to determine the
center of the peak positions with higher accuracy than classically possible, if
we are using the photon-number resolved peaks for $k\ge 3$. 
\section{Summary}
\par We have highlighted the difference between resolution and sensitivity in
interferometry measurements with classical and nonlcalssical sources and
 detectors. In particular we have shown that the resolution of a Fabry-P\'erot interferometer with
 weak coherent states and photon-number resolving detection is
 improved up to three times compared to a classical detection strategy.
 Here resolution relates to how well two peaks can be seen as distinct as
 opposed to the sensitivity in finding the center of one lone peak. The improvement in
 resolution is not to be confused with the sensitivity shown in Fig.~\ref{photon-number-resolved},
 which actually cannot be improved for the coherent state input and a
 photon-number resolving measurement. We also show that by replacing the
 classical input state with a photon-number state $|n\rangle$ incident on the
 FPI and performing a photon-number resolved measurement, we obtain
 supersensitivity (beat the shot-noise limit, Fig.~\ref{DeltaL}). A demonstration experiment to show this effect with photon pairs from a spontaneous
 parametric down conversion source or an optical parametric oscillator incident on a FPI should be well within reach. An interesting line of
 research would be to investigate generalized quantum metrology schemes other than
 the schemes above with
 coherent states and detection strategies based on photon-number resolving detectors.     
\begin{acknowledgments}
C.F.W. and J.P.D. acknowledge the Defense
Advanced Research Projects Agency Quantum Sensors Program. 
A.J.P., J.C., J.F., and A.M. acknowledge the MURI
Center for Photonic Quantum Information Systems ARO/IARPA, the IARPA Entangled
Source, and the Intelligence Community Postdoctoral Research Associateship
Programs for their support. We thank H. Lee,
N. Sauer, W. W. Johnson, S. Polyakov, and S. W. Nam for very helpful comments. C.F.W. thanks
all the members of the optical technology division at NIST, Gaithersburg for
their generous hospitality.\\ Approved for Public Release, Distribution Unlimited.
\end{acknowledgments}


\begin{thebibliography}{99}
\bibitem{Lincoln} D. Lincoln, Nucl. Inst. Meth. A {\bf 453}, 177 (2000).
\bibitem{Miller} A. J. Miller S. W. Nam, and J. M. Martinis, Appl. Phys. Lett. {\bf 83}, 791
  (2003).
\bibitem{Waks} E. Waks, E. Diamanti, B. C. Sanders, S. D. Bartlett, and Y. Yamamoto, Phys. Rev. Lett. {\bf 92}, 113602 (2004).
\bibitem{Appl} A. J. Miller, S. W. Nam, J. M. Martinis, and A. V. Sergienko,
  Appl. Phys. Lett. {\bf 83}, 791 (2003).
\bibitem{Woo} A. E. Lita, A. J. Miller, and S. W. Nam, Opt. Express {\bf 5}, 3032 (2008).
\bibitem{Lita} A. E. Lita, A. J. Miller, S. W. Nam, J. Low. Temp. Phys. {\bf
    151}, 125 (2008).
\bibitem{Mitchell} M. W. Mitchell, J. S. Lundeen, and A. M. Steinberg, Nature
  {\bf 429}, 161 (2004).
\bibitem{Ralph} P. P. Rohde, J. G. Webb, E. H. Huntington, and T. C. Ralph,
  New J. Phys. {\bf 9}, 233 (2007).
\bibitem{Dauler} E. A. Dauler, A. J. Kerman, B. S. Robinson, J. K. W. Yang,
  B. Voronov, G. Goltsman, S. A. Hamilton, and K. Berggren, J. Mod. Opt. {\bf
  56}, 364 (2009).
\bibitem{Caves} C. M . Caves, Phys. Rev. Lett. {\bf 45}, 75 (1980).
\bibitem{Resch} K. J. Resch, K. L. Pregnell, R. Prevedel, A. Gilchrist,
  G. J. Pryde, J. L. O'Brien, and A. G. White, Phys. Rev. Lett. {\bf 98}, 223601 (2007).
\bibitem{Hwang} H. Lee, P. Kok, and J. P. Dowling, J. Mod. Opt., {\bf 49},
  2325 (2002); J. P. Dowling, Contemp. Phys., {\bf 49}, 125 (2008).
\bibitem{Boto} A. N. Boto, P. Kok, D. S. Abrams, S. L. Braunstein, C. P. Williams,
  and J. P. Dowling, Phys. Rev. Lett., {\bf 85}, 2733 (2000).
\bibitem{Boyd} S. J. Bentley, and R. W. Boyd, Opt. Express {\bf 12}, 5735 (2004).
\bibitem{Shapiro} S.-H. Tan, B. I. Erkmen, V. Giovannetti, S. Guha, S. Lloyd,
  L. Maccone, S. Pirandola, and J. H. Shapiro, Phys. Rev. Lett. {\bf 101},
  253601 (2008).
\bibitem{Giovannetti} V. Giovannetti, S. Lloyd, L. Maccone, and J. H. Shapiro,
  Phys. Rev. A. {\bf 79}, 013827 (2009). 
\bibitem{Tsang} M. Tsang, J. H. Shapiro, and S. Lloyd, Phys. Rev. A {\bf 78},
  053820 (2008).
\bibitem{Lloyd} S. Lloyd, Science {\bf 321}, 1463 (2008).
\bibitem{Khoury} G. Khoury, H. S. Eisenberg, E. J. S. Fonseca, and D. Bouwmeester, Phys. Rev. Lett. {\bf 96}, 203601 (2006).
\bibitem{Huver} S. D. Huver, C. F. Wildfeuer, and J. P. Dowling, Phys. Rev. A
  {\bf 78}, 063828 (2008).
\bibitem{Agarwal} A. Kolkiran and G. S. Agarwal, Opt. Express {\bf 15}, 6798 (2007).
\bibitem{Kimble} H. J. Kimble et al., Phys. Rev. D {\bf 65}, 022002 (2002).
\bibitem{Loudon} M. Ley and R. Loudon, J. Mod. Opt. Vol. 34, No. 2, 227 (1987).
\bibitem{Glauber} R. J. Glauber, Phys. Rev. {\bf 131}, 2766 (1963).
\bibitem{Hwang2} H. Lee, U. Yurtsever, P. Kok, G. M. Hockney, C. Adami,
  S. L. Braunstein, and J. P. Dowling, J. Mod. Opt. {\bf 51}, 1517 (2004).
\bibitem{Dowling} J. P. Dowling, Phys. Rev. A. {\bf 57}, 4736 (1998).
\end{thebibliography}
\end{document}